# Strong X-ray Absorption in a BALQSO: PHL5200

Smita Mathur[1], Martin Elvis[1] and K. P. Singh[2,3]
1. Harvard-Smithsonian Center for Astrophysics,
60 Garden St, Cambridge MA 02138
2. Code 668, Lab. for High Energy Astrophysics,
NASA/GSFC, Greenbelt, MD 20771.
3. NRC Senior Research Associate, on leave from
Tata Institute of Fundamental Research, Bombay, India.
*Internet: smita@cfa.harvard.edu*

October 3, 1995

**Abstract**

We present *ASCA* observations of the z=1.98 prototype BALQSO: PHL5200. The source was detected in both SIS and GIS. A power-law spectrum ($\alpha_E = 0.6^{+0.9}_{-0.6}$) with large intrinsic absorption ($N_H = 1.3^{+2.3}_{-1.1} \times 10^{23}$ cm$^{-2}$) best describes the spectrum. Excess column density over the local Galactic value is required at the 99% confidence level. This detection suggests that, although BALQSOs are X-ray quiet, it is strong absorption in the BAL region that makes them appear faint to low energy X-ray experiments. The required intrinsic absorbing column density is two to three orders of magnitude larger than earlier estimates of column densities in BALQSOs. This implies that the BAL systems are much more highly ionized than previously thought.

# 1 Introduction

Associated absorption is common in the optical and ultraviolet spectra of quasars (Ulrich, 1988). A subset of these have very broad absorption line profiles extending up to $\Delta v = 0.1 - 0.2c$ redwards with respect to the quasar rest frame (see e.g. Turnshek 1988). These Broad Absorption Line quasars (BALQSOs), show absorption features due to high ionization lines of $C^{+3}$, $Si^{+3}$, and other ions. Low ionization BALQSOs have also been observed which show $Mg^{+1}$ and/or $Al^{+2}$ absorption troughs. BALQSOs have been estimated to have column densities $N_H \sim 10^{20-21}$ cm$^{-2}$ (Turnshek 1984, Hamann *et al.* 1993). As a class, BALQSOs share some common properties: they are always radio-quiet (Stocke et al. 1992), may have abundances 10-100 times solar (in their emission lines; Turnshek 1988, Hamann & Ferland 1993) and are X-ray quiet (Green *et al.* 1995). Recent work suggests that BALQSOs are normal radio-quiet quasars seen from an unusual direction (Weymann *et al.* 1991, Hamann *et al.* 1993). In this case all radio-quiet quasars have collimated BAL outflows, which however are pointed out of our line of sight in some 90% of cases. Thus BALQSOs, far from being exotic objects, give us a special probe into the gas dynamics around the typical quasar.

However, physical conditions in the absorbing gas in the BALQSOs are poorly determined from optical/UV absorption line studies (Lanzetta *et al.* 1991). This is because only a few, usually saturated, lines are measured yielding lower limits to column densities for a few ions, but little information on the ionization state. If, as in the narrow line associated absorbers, there is X-ray absorption as well as optical and UV, then the combined X-ray and UV analysis would allow us to derive the physical conditions in BALQSO absorption systems (Mathur *et al.* 1994, Mathur 1994, Mathur *et al.* 1995). This, however, has been difficult, since BALQSOs are elusive X-ray sources, and so are otherwise essentially unconstrained in their X-ray properties. In a soft X-ray study of quasars with *EINSTEIN* only four out of nine BALQSOs were detected (Zamorani *et al.* 1981). Initial results from *ROSAT* are mainly upper limits (Kopko *et al.* 1993, Green *et al.* 1995) implying that they are relatively faint in soft X-rays (i.e. have steep $\alpha_{ox}$). Our understanding of BALQSOs is incomplete without knowing their X-ray properties. In fact, lack of knowledge of the underlying ionizing continuum is one of the major uncertainties in the models of BALQSOs: Are they intrinsically X-ray quiet (i.e. large $\alpha_{ox}$)? Or, is it strong absorption that makes them look faint?

PHL5200, a prototype BALQSO at z=1.98 (Burbidge, 1968), was detected in hard X-rays by the *EXOSAT* medium energy (ME) experiment but not by the low energy (LE) experiment (Singh *et al.*1987). To obtain consistency between the *EXOSAT* ME and LE requires a column density of $\geq 10^{22} atoms\ cm^{-2}$ making it an excellent candidate for examining the BAL region. We observed PHL5200 with *ASCA* with this aim in mind. *EINSTEIN* did not detect (Zamorani *et al.* 1981) and *ROSAT* has not observed PHL5200.

## 2  *ASCA* Observations and Data Analysis

*ASCA* (Tanaka *et al.* 1994) observed PHL5200 on 1994 June 21 for a net exposure time of 17.7 ksec (Table 1). *ASCA* has two Solid-state Imaging Spectrometers (SIS0 and SIS1, Loewenstein & Isobe, 1992) and two Gas Imaging Spectrometers (GIS3 and GIS4, Ohashi *et al.* 1991). The SIS were operated in 2-CCD mode (see figure 1). The source was faint but was clearly detected in SIS0 and GIS3 (figure 1). The X-ray position from SIS0 is (J2000) 22:28:26, −5:18:54; 1.1 arcminute from the optical position (Schneider *et al.* 1992), consistent with the current satellite pointing uncertainties (Tanaka *et al.* 1994). The source was off axis in SIS1 (where it lay close to the gap between two chips) and GIS2 and was not detected in either. This is consistent with the fact that the optical axes of telescopes containing SIS0 and GIS3 are much closer to each other than the others. No other sources were seen in any of the instruments to a level similar to the count rates of PHL5200.

Data were extracted in standard way using the FTOOLS and XSELECT software. [1] Standard screening criteria were used as recommended in *ASCA* ABC guide: >10 degree bright earth angle; and a cut-off rigidity of >6 GeV/c. Hot and flickering pixels were removed from the SIS data using XSELECT. All SIS events of grade 0, 2, 3 and 4 were accepted. Data of both faint and bright modes with high, medium and low telemetry rates were combined. These data can be combined without any calibration compromises. *ASCA* X-ray telescope has a broad point spread function and jittering of the spacecraft can appear on arcminute scales. To take this into account, source counts were extracted from a circular region of 6 arcminute radius for GIS3 and from a 4 arcminute radius for SIS0. The source was pointed at the center of chip # 1 of SIS0, putting the bulk of its photons into just one chip. Detectors SIS0 and GIS3 yielded $\sim$ 500 total counts each. Data from these detectors cannot be combined since they have different properties. The background was estimated using the same spatial filter on the deep field background images (*ASCA* ABC guide). A background subtracted count rate of $(1.04 \pm 0.15) \times 10^{-2}$ was observed by SIS0, and $(6.26 \pm 1.47) \times 10^{-3}$ by GIS3. The data were grouped to contain at least 10 counts (background subtracted) per pulse height analysis (PHA) channel to allow the use of the Gaussian statistic. The data have modest signal to noise ratio; however it can be clearly seen that that there are essentially no counts below $\sim$ 1 keV ($\sim$ 3 keV in the rest frame)(see Figure 2). The highest rest energy detected for PHL5200 is 12 keV at $3\sigma$ for 0.5 keV wide bins. Figure 2 shows the SIS and GIS spectra of PHL5200.

---

[1] FTOOLS is a collection of utility programs to create, examine or modify data files in FITS format. XSELECT is a command line interface to the FTOOLS, for X-ray astrophysical analysis. The software is distributed by the ASCA Guest Observer Facility.

The SIS and GIS spectra extracted in this way were then analyzed using XSPEC. The March 1995 release of the response matrices was used for the GIS data, and the November 1994 release for the SIS data. A power law spectrum with fixed Galactic absorption ($4.8 \times 10^{20}$ atoms cm$^{-2}$; Stark *et al.* 1992) provides an acceptable fit to the SIS0 data (Table 2). However, if absorption is allowed to be a free parameter, then the fit is improved with >98% confidence (F-test, Table 2). The fitted value ($N_H(z=0) = 9 \times 10^{21}$ atoms cm$^{-2}$, solar abundance) is much larger than the Galactic column density towards PHL5200, indicating excess absorption along the line of sight. This is also much larger than the uncertainties in the SIS low energy response, which may overestimate the column density by up to $2 \times 10^{20}$ cm$^{-2}$ (Day, C. S. R. 1995. "Calibration Uncertainties", *ASCA* GOF WWW page. URL: http://heasarc.gsfc.nasa.gov/docs/asca/cal_probs.html). We then fitted a power-law spectrum with Galactic column and an additional column of absorber allowing its redshift to be free. We found no preferred redshift for the additional absorber. Fixing the absorber at the source gives a column density of $1.4^{+2.0}_{-1.2} \times 10^{23}$ cm$^{-2}$ (90% confidence for one parameter, solar abundance). The power-law energy index is $\alpha_E = 0.8^{+1.1}_{-0.9}$.

For the GIS data, a similar fit of a power-law spectrum with fixed Galactic and additional z=1.98 absorption is acceptable, but does not constrain the parameters well because the data have large errors (Table 2).

A combined SIS and GIS analysis does constrain the parameters of the model slightly better (Figure 3, Table 2). The column density at the source is $1.3^{+2.3}_{-1.1} \times 10^{23}$ cm$^{-2}$ and $\alpha_E = 0.6^{+0.9}_{-0.6}$. This excess absorption, above Galactic N$_H$, is required at 99% confidence (F-test).

The 2–10 keV (observed frame) flux is $2.9^{+13.9}_{-1.3} \times 10^{-13}$ ergs s$^{-1}$ cm$^{-2}$ (corrected for best fit absorption) and a 2–10 keV (rest frame) luminosity is $9.3 \times 10^{45}$ ergs s$^{-1}$ (H$_0$ = 50, q$_0$ = 0). The flux in the EXOSAT ME band (2–6 keV observed) is $2^{+6}_{-1} \times 10^{-13}$ ergs s$^{-1}$ cm$^{-2}$. This is smaller than the *EXOSAT* flux ($\sim 2 \times 10^{-12}$ ergs s$^{-1}$ cm$^{-2}$; Singh *et al.* 1987) by at least a factor of 2.5. The optical continuum of PHL5200 does not vary by such a large amount (Barbieri *et al.* 1978). It is possible that it is variable absorption rather than intrinsic source variability that might be responsible for the difference in the *ASCA* and *EXOSAT* ME fluxes. The *ASCA* flux is consistent with the upper limits observed by the EINSTEIN IPC ($< 4.5 \times 10^{-13}$ ergs s$^{-1}$ cm$^{-2}$) and the *EXOSAT* CMA ($< 5 \times 10^{-13}$ ergs s$^{-1}$ cm$^{-2}$).

The *ASCA* derived monochromatic luminosity at 2 keV (rest frame) is $1.3 \times 10^{28}$ ergs s$^{-1}$ Hz$^{-1}$ and at 2500 Å (rest frame) it is $1.2 \times 10^{32}$ ergs s$^{-1}$ Hz$^{-1}$ (Zamorani *et al.* 1981), giving $\alpha_{ox} = 1.5$.

An Fe-K absorption edge is not detected ($\tau < 0.9$, 90% confidence for one interesting parameter). The opacity of an Fe edge corresponding to $N_H = 10^{23}$ cm$^{-2}$ is $\tau = 0.1 f_{ion}$ where $f_{ion}$ is the ionization fraction of iron in hydrogen-like state. Our data are not sensitive enough to detect such an edge.

The Fe-K emission line (Ross & Fabian 1993) is also not detected (see figure 2). We place a 0.5 keV upper limit (90% confidence for one interesting parameter) to the rest frame equivalent width of a narrow ($\sigma < 10$ eV) line between 2.1 and 2.4 keV (6.3–7.1 keV rest frame). This can be used to place an upper limit on the covering factor of the absorber. If the absorber is a uniform spherical shell surrounding the X-ray continuum source, then the Fe K$\alpha$ line flux through recombination after photoionization of helium-like iron is given by I$_{line}$=(N$_H A(Fe)/10^{19.8})(\Omega/4\pi)$ I$_{abs}\eta$ (Basko 1980). $\eta$ is the fluorescent yield, the efficiency with which the flux above 7.1 keV (I$_{abs}$) is re-emitted as an Fe-K line. Assuming solar abundance of iron (A(Fe)=3.3×10$^{-5}$, Grevesse & Andres 1989) and $\eta = 0.5$ (Krolik & Kallman, 1987) we estimate the covering factor of the line emitting region, $\Omega/4\pi < 4 f_{ion}^{-1}$; which is not an interesting limit. If, however, the heavy element abundance is 10 times solar (Hamann & Ferland 1993) then $\Omega/4\pi < 0.4\ f_{ion}^{-1}$, consistent with Hamann

*et al.* (1993).

# 3 Discussion

The *ASCA* spectrum of PHL5200 shows excess absorption at 99% confidence. A column density of $0.2 - 4 \times 10^{23} \frac{Z_\odot}{Z}$ cm$^{-2}$ is obtained if the absorber is at the source. A power-law was a good fit to the data with the spectral slope ($\alpha_E = 0.6^{+0.9}_{-0.6}$) in the normal range (Wilkes *et al.* 1994). The PHL5200 value of $\alpha_{ox} = 1.5$, is also normal for a radio quiet quasar (Wilkes *et al.* 1994).

The inferred absorbing column density for PHL5200 is two to three orders of magnitude larger than the earlier estimates of column density in BALQSOs (Hamann *et al.* 1993, Turnshek 1984). This implies that the BAL clouds may be more highly ionized ($N_{HI}/N_H \sim 10^{-8}$) than previously thought ($N_{HI}/N_H \sim 10^{-5}$, Hamann *et al.* 1993), as was true with narrow associated absorbers (Mathur *et al.* 1994, 1995). The estimates from the saturated UV lines appear to have been misleading. Recent models of BALQSOs (Murray *et al.* 1995), however, do consider column densities as large as we find in PHL5200. If, on the other hand the abundences are 100 – 1000 times solar then the Hydrogen column density would be smaller ($N_H \sim 10^{20}$ cm$^{-2}$). However, the ionization state would still be high, since the comparison is between metal line absorption in the UV and absorption in the X-ray. The column density in PHL5200 is also about an order of magnitude larger than other, narrow, associated absorption systems (Fiore *et al.* 1993, Turner *et al.* 1994). In this respect, as they are in velocity width, the BALQSOs may be extreme examples of other associated absorbers.

This is consistent with our earlier conjecture that all associated absorbers may form a continuum of properties with column density, outflow velocity and the distance from the central continuum (Mathur *et al.* 1994). Are BALQSOs also similar to these in being 'XUV absorbers'? i.e. are the broad absorption lines observed in the UV caused by the same matter producing X-ray absorption? This can be investigated by combined analysis of X-ray and UV spectra (Mathur *et al.* 1994, 1995) of PHL5200; but is beyond the scope of this paper. If they are indeed the same, it would allow us to further constrain the physical properties of the absorber and so of the outflowing circumnuclear matter (Mathur *et al.* 1995).

The present study implies that BALQSOs are not intrinsically X-ray quiet; it is the extreme absorption that makes them appear faint to low energy experiments. Since the absorption is significant only in soft X-rays, hard X-ray observations, above a few keV, would reveal their presence as X-ray sources. This can be done with missions like *ASCA*, XTE, SAX and AXAF. We have been awarded XTE time to observe BALQSOs with this aim.

**Acknowledgements:**

This research has made use of the NASA/IPAC Extragalactic Database (NED) which is operated by the Jet Propulsion Laboratory, CALTECH, under contract with the National Aeronautics and Space Administration. This work was supported by NASA grants NAGW-2201 (LTSA), NAG5-2563 (ASCA), NAGW-4490 (LTSA) and NASA contract NAS8-39073 (ASC).

Table 1: ASCA Observations of PHL5200

| Instrument | Total Counts | Exposure (s) | Net Count Rate $s^{-1}$ |
|---|---|---|---|
| SIS0 | 513 | 16587 | $0.01\pm0.001$ |
| GIS3 | 505 | 16788 | $0.006\pm0.001$ |

Table 2: Spectral fits to ASCA data of PHL5200

| Data | Model | $\alpha_E$ | $N_H$ (free)[a] | Normalization[b] | $\chi^2$ (dof)[c] |
|---|---|---|---|---|---|
| SIS | Power-law: | | | | |
|  | $+N_H$ | $0.9^{+1.3}_{-1.0}$ | $0.9^{+1.2}_{-0.8}$ | $1.3^{+4.8}_{-0.3}$ | 5.2 (12) |
|  | $+N_H$ (Gal.) fixed | $-0.1^{+0.4}_{-0.4}$ | | $0.3^{+0.1}_{-0.2}$ | 8.4 (13) |
|  | $+N_H$ (z=1.98) | $0.8^{+1.1}_{-0.9}$ | $14.0^{+19.7}_{-12.4}$ | $1.2^{+3.4}_{-0.4}$ | 4.8 (12) |
| GIS | Power-law: | | | | |
|  | $+N_H$ | $2.0^{+3.1}_{-1.8}$ | $4.5^{+0.0}_{-3.8}$ | $9.6^{+875}_{-0.5}$ | 5.4 (9) |
|  | $+N_H$ (Gal.) fixed | $-0.1^{+0.5}_{-0.6}$ | | $0.3^{+0.2}_{-0.1}$ | 9.8 (10) |
|  | $+N_H$ (z=1.98) | $2.8^{+6.2}_{-2.6}$ | $130^{+0}_{-118}$ | $47^{+170}_{-0.5}$ | 5.7 (9) |
| SIS+GIS | Power-law: | | | | |
|  | $+N_H$ | $0.6^{+0.0}_{-0.7}$ | $0.9^{+1.4}_{-0.7}$ | $1.0^{+2.9}_{-0.4}$ | 14.1 (24) |
|  | $+N_H$ (Gal.) fixed | $-0.1^{+0.3}_{-0.3}$ | | $0.3^{+0.1}_{-0.2}$ | 18.4 (25) |
|  | $+N_H$ (z=1.98) | $0.6^{+0.9}_{-0.6}$ | $13.1^{+23.2}_{-11.1}$ | $0.9^{+3.0}_{-0.4}$ | 14.2 (24) |

a: $\times 10^{22}$ cm$^{-2}$

b: in units of $10^{-4}$ photons keV$^{-1}$ cm$^{-2}$ s$^{-1}$ at 1 keV

c: degrees of freedom.

**Figure Captions:**
**Figure 1:** ASCA GIS3 (left) and SIS0 (right) grey scale images around PHL5200. North is 66.7 degrees clockwise from the top. The GIS field of view is 50 arcmin diameter, and each SIS chip is 11.1 arcmin on a side (ASCA Technical Description, 1993)

**Figure 2:** ASCA spectral data (crosses) with best fit power law with fixed Galactic and intrinsic absorption models: SIS (top), GIS (middle), Both SIS & GIS (bottom).

**Figure 3:** Confidence contours for the combined SIS and GIS spectrum. Contours of 68%, 90% and 99% confidence regions are shown. The Galactic column density is shown as a dashed line.

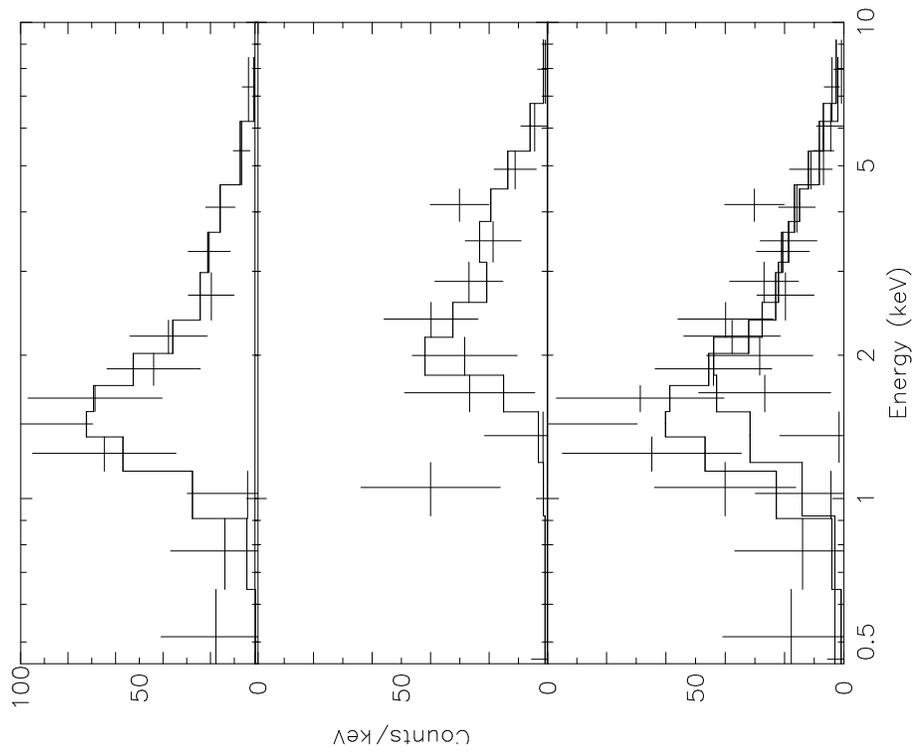

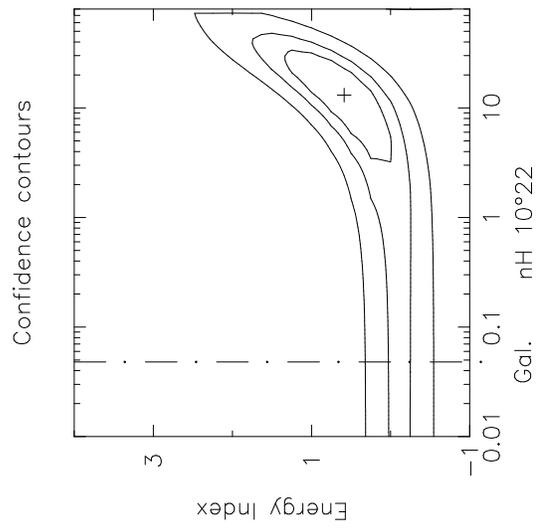